\documentclass[journal]{IEEEtran}
\usepackage{siunitx}
\usepackage{amsfonts,amsmath,amsthm,booktabs}
\usepackage{graphicx,xspace,epsfig,syntonly,dsfont}
\usepackage{url,cite,bm,bbm}
\usepackage{algorithmicx}
\usepackage{multirow, makecell}
\usepackage{arydshln}
\usepackage{booktabs}
\usepackage{hhline}
\usepackage{balance}
\usepackage{empheq}
\usepackage[caption=false, font=footnotesize]{subfig}
\usepackage{stfloats}
\usepackage{authblk}
\usepackage{diagbox}
\usepackage{multirow}
\usepackage{enumitem}
\newcommand{\xmark}{\ding{55}}
\usepackage{pifont}

\usepackage{algorithm}
\usepackage{algpseudocode}

\long\def\symbolfootnote[#1]#2{\begingroup
\def\thefootnote{\fnsymbol{footnote}}
\footnote[#1]{#2}\endgroup}
\DeclareSIUnit \voltampere {MVA} 

\IEEEoverridecommandlockouts

\allowdisplaybreaks[4]

\title{A hard-constrained NN learning framework for rapidly restoring AC-OPF from DC-OPF}
\author{
    \IEEEauthorblockN{
        Kejun Chen\IEEEauthorrefmark{1},
        Bernard Knueven\IEEEauthorrefmark{1},
        Wesley Jones\IEEEauthorrefmark{1}
    }
    \thanks{\IEEEauthorblockA{\IEEEauthorrefmark{1}Computational Science Center, National Laboratory of the Rockies (NLR), Golden, CO, USA.} Emails: \texttt\{kejun.chen,  bernard.knueven, wesley.jones\}@nlr.gov}

    \thanks{
    This work was authored by the National Laboratory of the Rockies for the U.S. Department of Energy (DOE), operated under Contract No. DE-AC36-08GO28308. This work was supported by the Laboratory Directed Research and Development (LDRD) Program and Modeling, Simulation, Optimization Capability for Grid Modernization, Energy Systems Integration, Mobility Electrification (MSOC 4 GEM) at the National Laboratory of the Rockies. The views expressed in the article do not necessarily represent the views of the DOE or the U.S. Government. The U.S. Government retains and the publisher, by accepting the article for publication, acknowledges that the U.S. Government retains a nonexclusive, paid-up, irrevocable, worldwide license to publish or reproduce the published form of this work, or allow others to do so, for U.S. Government purposes. }
    
    \thanks{This research was performed using computational resources sponsored by the U.S. Department of Energy's Office of Critical Minerals and Energy Innovation and located at the National Laboratory of the Rockies.}
}

\begin{document}
\maketitle
\begin{abstract}
This paper proposes a hard-constrained unsupervised learning framework for rapidly solving the non-linear and non-convex AC optimal power flow (AC-OPF) problem in real-time operation. Without requiring ground-truth AC-OPF solutions, feasibility and optimality are ensured through a properly designed learning environment and training loss. Inspired by residual learning, the neural network (NN) learns the correction mapping from the DC-OPF solution to the active power setpoints of the generators through re-dispatch. A subsequent optimization model is utilized to restore the optimal AC-OPF solution, and the resulting projection difference is employed as the training loss. A replay buffer is utilized to enhance learning efficiency by fully leveraging past data pairs. The optimization model is cast as a differentiable optimization layer, where the gradient is derived by applying the implicit function theorem to the KKT conditions at the optimal solution. Tested on IEEE-118 and PEGASE-9241 bus systems, numerical results demonstrate that the proposed NN can obtain strictly feasible and near-optimal solutions with reduced computational time compared to conventional optimization solvers. In addition, aided by the updated DC-OPF solution under varying topologies, the trained NN, together with the PF solver, can rapidly find the corresponding AC solution. The proposed method achieves a $40\times$ time speedup, while maintaining an average constraint violation on the order of $10^{-4}$ and an optimization gap below $1\%$.

\end{abstract}

\begin{IEEEkeywords}
Optimal power flow, hard-constrained neural network, differential optimization layer, hybrid learning framework. 
\end{IEEEkeywords} 

\begin{section}{Introduction}
Optimal power flow (OPF) plays an essential role in the power grid operation and planning. It aims to find the best operating point that minimizes the generation cost while satisfying a set of operational constraints. The physical constraints include the power flow (PF) balance equations and the operating limits of voltage magnitudes, power generation, and branch flows. Due to the highly non-linear AC-PF equations, the AC-OPF problem is inherently non-convex and non-linear, which makes it an NP-hard problem \cite{Karsten2016}. Thus, the computation burden can be heavy for a non-linear programming solver to find a local solution for the large-scale bus system. Moreover, the independent system operator may need to solve the AC-OPF problem more frequently in the real-time market due to the stochastic load demand and power generation output \cite{Chatzos2021}. Purely relying on the conventional optimization solvers cannot handle this new challenge due to the uncertainty in the performance of optimization solvers for non-convex non-linear problems. In practice, the low-fidelity OPF model is adopted in the real-time market, so-called DC-OPF. However, the DC-OPF solution can never be AC-feasible \cite{Baker2021}. Thus, there is an increasing interest in developing rapid AC-OPF solvers. 

\begin{subsection}{Related work}
Impressed by the universal approximation capabilities of neural networks (NNs), deep learning techniques have been widely used to solve constrained optimization problems. In AC-OPF, the NN can obtain a highly feasible and near-optimal solution rapidly, which makes it appealing in real-time operation \cite{Babiker2025}. Depending on the reliance on conventional optimization solvers, end-to-end learning frameworks are divided into supervised and unsupervised learning categories. 

In supervised learning, the NN is trained to predict the optimal solution based on the historical data pairs generated by conventional optimization solvers \cite{Fioretto2020, Chatzos2021, Lei2021}. The NN weights are updated to minimize the mean square error between the NN outputs and the targets. However, it is not a trivial task to generate an extensive training dataset for large-scale systems due to the high computational cost. 

Increasing attention has focused on developing unsupervised learning frameworks to eliminate the need for ground truth labels. Ref. \cite{Wanjun2024} utilizes the NN to predict the voltage magnitudes and phase angles and penalizes the load mismatch after reconstruction. However, in reality, the voltage phasors of the load buses are not controllable for the independent system operator. Additionally, this learning framework design can lead to significant load mismatches in the PF equations. To ensure the strict satisfaction of the equality constraints, prior work~\cite{Pan2020DeepOPFAF, donti2021dc3, Kejun2024} utilized the NN to predict a set of controllable variables and recover the remaining dependent variables. The PF solver embedded within the learning framework can also eliminate load mismatches. 

However, the integration of the PF solver into the training framework introduces some new challenges, such as the PF solver divergence issue. If the subsequent PF solver cannot find any feasible solution, this can result in a training failure. This occurs because the randomly initialized NN may generate infeasible decision variable values. The divergence issue occurs frequently on large-scale systems because the NN must search over an extensive feasible space of decision variables. Ref.~\cite{Jiayu2024} utilizes an optimization model that projects the infeasible generator setup values to the feasible ones and verifies the effectiveness on a small bus system. However, when tested on the large-scale bus system, the NN cannot generate a highly AC-feasible solution directly, which implies the optimization model is necessary to ensure feasibility even during the testing phase. This results in a longer computational time compared to using the PF solver. To address the divergence issue in the initial training stage, refs. \cite{WANG2022} and \cite{Kejun2025} use the warm-up strategy based on the AC-OPF solutions to provide good initial weights for the NN. The NN weights are obtained by minimizing the errors of NN outputs and the ground truth labels generated by conventional optimization solvers. Refs.~\cite{Wenbo2024} and \cite{Seonho2025} utilize a power balance repair layer for the economic dispatch problem and the DC-OPF problem, respectively, while the heuristic balance tuning does not work well for the AC-OPF problem due to line loss.

In this work, we mitigate the solver divergence issue by incorporating the optimization-based feasibility restoration model. The optimization model projects the infeasible solution, which comprises the NN residual correction and the DC-OPF solution, to the closest AC-feasible point. The projection cost brought about by the feasibility restoration is minimized in the training loss function, which helps guide the NN updates toward feasible regions. Inspired by residual learning, ref.~\cite{Kejun2023} shows that pushing the residual correction regarding the linear relationship to zero is more straightforward than learning the original complicated non-linear mapping \cite{he2015}. Thus, we utilize the DC-OPF solution to provide a good reference value for the NN to learn the correction to the AC-solution. The DC-OPF solution is generated using the conventional optimization solver, but it requires much less running time than generating the ground truth AC-OPF solutions. 

Without explicit labels, the NN training in the unsupervised learning framework relies on a properly designed loss function, as it will play a crucial role in ensuring feasibility and optimality. It typically consists of the operational constraint violation penalty and the generation cost. However, one of the challenges for multi-task learning is tuning the weighting parameters for multiple loss terms, which often requires an exhaustive search \cite{Groenendijk2021}. Multiple loss items may lead to gradient pathologies, and the dominant loss may cause the NN to update primarily towards one certain direction \cite{Wanjun2024, Sifan2021, Shaohui2023}. In the proposed framework, the optimization model ensures feasibility, thereby alleviating the challenges associated with multi-objective training.

In addition, this paper shows the capability of NN in solving optimization problems with non-linear hard constraints. The embedded optimization model can be treated as a differentiable optimization layer. Ref. \cite{tordesillas2023, min2025, amos2017} propose an NN learning framework with hard convex constraints. Ref. \cite{ferber2019} proposes an approach to differentiating through disciplined convex programs based on CVXPY \cite{diamond2016cvxpy}. However, they are not readily applicable to the AC-OPF problem due to the non-convex and non-linear operational constraints. We propose a differentiable optimization layer based on Pyomo \cite{bynum2021pyomo}, where the forward mapping adopts a nonlinear programming solver such as \emph{Ipopt} to obtain the optimal solution. In the backpropagation, the gradient of the optimal solution w.r.t. the input can be derived by applying the implicit function theorem to the KKT conditions \cite{duvenaud2023, Pirnay2012, Jiayu2024}. 

Additionally, it is essential to ensure safe system operation under topology changes. Refs.\cite{GIRAUD2024, Seonho2025} utilize line outage distribution factors to redistribute the power flows when contingencies happen. Compared to the fully connected neural network (FCNN), graph neural networks can effectively handle network topology changes through neighborhood aggregation \cite{lovett2024opf}. In this work, we utilize an FCNN to learn the mapping from the DC-OPF solution to the AC-OPF solution. The topology changes can be reflected in the DC-OPF solution and implicitly represented in the FCNN input. This pivots the topology change to the generalization ability of the FCNN in learning the DC to AC mapping. The contributions of this paper can be summarized as:
\begin{itemize}
    \item A hard-constrained NN learning framework is proposed to rapidly obtain a highly feasible and near-optimal AC-OPF solution under varying topologies. 
    
    \item The proposed method does not rely on conventional solvers to generate explicit ground truth AC-OPF targets. Given that the DC solution may not be AC-infeasible, the embedded optimization layer finds the closest optimal AC-OPF solution for the NN to train. 
    
    \item Inspired by residual learning, the NN learns the residual correction of the optimal power setpoints to the DC-OPF solution. A replay buffer is adopted to enhance training efficiency by utilizing past data pairs.

\end{itemize}
The remainder of this paper is organized as follows. Section~\ref{sec:prob} formulates the AC-OPF problem. Section~\ref{sec:method} presents the proposed hybrid learning framework with a differentiable optimization layer. Section~\ref{sec:results} evaluates its performance on two benchmark systems. Finally, Section~\ref{sec:conclusion} offers the conclusion.

\end{subsection}

\end{section}

\begin{section}{Problem formulation}\label{sec:prob}
Consider an electrical grid with $N$ buses (consisting of generator buses $\mathcal{N}_g$, load buses $\mathcal{N}_d$, and one reference bus) and $\mathcal{M}$ transmission lines. The AC-OPF problem minimizes the total generation cost while satisfying a set of operational constraints \cite{Cain12}:
\begin{subequations}
\label{AC-PF}
\begin{align} 
    &\min_{\mathbf{V}, \boldsymbol{\theta}, \mathbf{P}_g, \mathbf{Q}_g, \mathbf{P}_{ij}, \mathbf{Q}_{ij}} \quad  \sum_{i} c_i(P_{g, i}) \label{oj}\\
    &\textrm{s.t.} \quad P_{g, i} - P_{d, i} = V_i \sum_{j=1}^N V_j (G_{ij}\cos \theta_{ij} + B_{ij}\sin \theta_{ij} )  \label{pf1} \\ 
    & Q_{g, i} - Q_{d, i} = V_i \sum_{j=1}^N V_j (G_{ij}\sin \theta_{ij} - B_{ij}\cos \theta_{ij} )  \label{pf2}\\ 
    & P_{ij} = -G_{ij}V_i^2 + V_iV_j(G_{ij}\cos \theta_{ij} + B_{ij}\sin \theta_{ij})  \label{bf1}\\
    & Q_{ij} = B_{ij}V_i^2 + V_iV_j(G_{ij}\sin \theta_{ij} - B_{ij}\cos \theta_{ij})  \label{bf2} \\ 
    & P_{ij}^2 + Q_{ij}^2 = |S_{ij}|^2\, , \forall  (i, j)\in \mathcal{M}  \label{bf3} \\
    & |S_{ij}|^2 \leq (S_{ij}^{\text{max}})^2,\, \forall  (i, j)\in \mathcal{M}  \label{bf4} \\
    & \underline{V}_{i} \leq V_{i} \leq \overline{V}_{i},\, \forall  i\in \mathcal{N} \label{v} \\ 
    & P_{g, i} = Q_{g, i} = 0 \, , \forall  i \in \mathcal{N}_d \label{no_gen} \\
    & \theta_{\text{ref}} = 0  \label{ang} \\
    & \underline{P}_{g, \text{ref}} \leq P_{g, \text{ref}} \leq \overline{P}_{g, \text{ref}}\, , \,
    \underline{Q}_{g, \text{ref}} \leq Q_{g, \text{ref}} \leq \overline{Q}_{g, \text{ref}}\, \label{pgqg_ref} \\
    & \underline{Q}_{g, i} \leq Q_{g, i} \leq \overline{Q}_{g, i}\, , \forall  i\in \mathcal{N}_g   \label{qg}\\
    & \underline{P}_{g, i} \leq P_{g, i} \leq \overline{P}_{g, i}\, , \forall  i\in \mathcal{N}_g. \label{pg}
\end{align} 
\end{subequations}
Let $P_{d, i}/Q_{d, i}$ and $P_{g, i}/Q_{g, i}$ represent the active/reactive load demands and power generation outputs at bus $i$. $P_{ij}$, $Q_{ij}$, and $S_{ij}$ represent the active, reactive, and apparent power flows from bus $i$ to bus $j$. $V_i$ denotes the voltage magnitude of bus $i$. $\theta_{ij} := \theta_i - \theta_j$ denotes the phase angle difference between bus $i$ and $j$. Let $G_{ij}$ and $B_{ij}$ denote the real and imaginary parts of the $(i,j)$-th element of the nodal admittance matrix, respectively. The subscript $(\cdot)_\text{ref}$ denotes the corresponding decision variable of the reference bus. In addition, the objective function~\eqref{oj} describes the total generation cost, where $c_i(\cdot)$ denotes the cost function of generator $i$. Eqs.~\eqref{pf1} and \eqref{pf2} are the AC-PF balance equations. Eqs.~\eqref{bf1}, \eqref{bf2}, and \eqref{bf3} are the active, reactive, and apparent branch flow equations, respectively. The upper bounds of the apparent power flows are given in Eq.~\eqref{bf4}. Eq.~\eqref{v} describes the operating limits of voltage magnitudes. Eq.~\eqref{no_gen} implies that no generators are connected to the load bus. Eq.~\eqref{ang} implies the phase angle of the reference bus is fixed at 0. Eq.~\eqref{pgqg_ref} depicts the operating limits of the active and reactive power generation outputs of the reference bus. Eqs.~\eqref{qg}-\eqref{pg} describe the operating limits of the active and reactive power outputs of generators. 
\end{section}

\begin{section}{Proposed hybrid learning framework}\label{sec:method}
\begin{subsection}{Variable splitting}
Fig.\ref{framework} depicts the proposed learning framework. Let $\mathbf{x}:= [\mathbf{P}_d; \mathbf{Q}_d] \in \mathbf{R}^{2N}$ collect the load demands of all buses. The voltage magnitudes of all the buses of the DC-OPF solution are $1$, so they are not used as the NN input features. $\mathbf{P}_g^{\text{DC}} \in \mathbf{R}^{N_g}$ denotes the active power generations of the DC-OPF solution and $\Delta\mathbf{P}_g^{\text{AC}} \in \mathbf{R}^{N_g} $ represents the corresponding residual correction to the AC-OPF setpoints. Let $\mathbf{y}:= [\mathbf{P}_g; \mathbf{V}_g; V_\text{ref}] \in \mathbf{R}^{2N_g + 1}$ collect the active power outputs of the generator buses, the voltage magnitudes of the generator buses, and the voltage magnitude of the reference bus. 

To ensure the strict satisfaction of PF equations, Ref.~\cite{donti2021dc3} employs the NN to predict a subset of decision variables and utilizes a follow-up PF solver to recover the remaining uncontrollable decision variables. Specifically, given $\mathbf{x}$ and $\mathbf{y}$, the unknown decision variables, including voltage magnitudes of the load buses $\mathbf{V}_d$, the phase angles of the load buses $\boldsymbol{\theta}_d$, and the phase angles of generator buses $\boldsymbol{\theta}_g$ can be calculated by solving the Eqs.$~\eqref{pf1}_{\mathcal{N}_d \cup \mathcal{N}_g}$ and $~\eqref{pf2}_{\mathcal{N}_d}$. After obtaining the voltage phasors of all the buses, the other remaining decision variables, such as reactive power generation outputs and branch flows, can be determined by solving the remaining equality constraints. We adopt this variable-splitting strategy to alleviate the NN training difficulty by predicting only a small subset of controllable variables.

 \begin{figure}[ht!]
    \centering
    \includegraphics[width=1\linewidth]{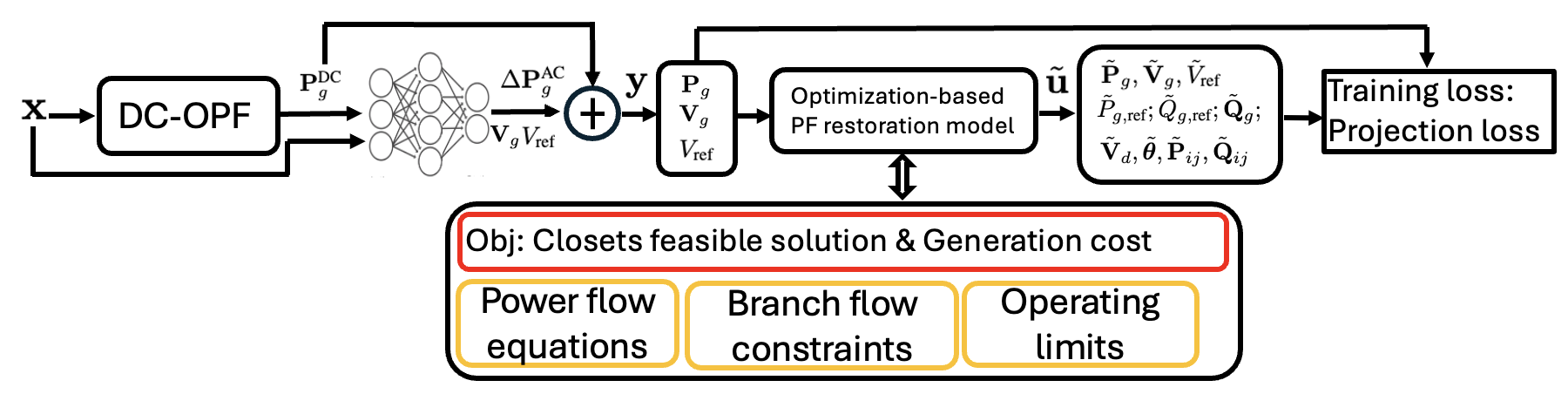}
    \caption{The proposed optimization-based end-to-end unsupervised learning framework for AC-OPF.}
    \label{framework}
\end{figure}

Besides voltage magnitudes, the proposed NN learns the residual correction $\Delta \mathbf{P}_g^\text{AC}$ of the active power setpoints of the generators. Since the NN weights are typically randomly initialized with small weights, the NN output values in $\Delta \mathbf{P}_g^\text{AC}$ are close to zero in the initial training stage. This naturally leads to $\mathbf{y} \approx \mathbf{P}_g^\text{DC}$, which in turn contributes to minimizing the deviation from the DC-OPF solution. However, the NN output $\mathbf{y}$ can be AC-infeasible due to improper setup values of voltage magnitudes and active power generation outputs. In this work, we utilize an optimization-based PF feasibility restoration model to project the infeasible decision variables to the closest optimal solution that satisfies all the operational constraints.

\end{subsection}

\begin{subsection}{The proposed optimization-based learning model}
The FCNN is adopted to learn the optimal mapping $\{ \mathbf{x}, \mathbf{P}_g^{\text{DC}} \} \mapsto \{ \Delta\mathbf{P}_g^{\text{AC}}, \mathbf{V}_g, V_{\text{ref}}\} $. Let $\mathbf{W}_{l}$ and $\mathbf{b}_{l}$ denote the weight matrix and bias vector of the $l$-th layer of the FCNN. The forward propagation can be expressed as:
\begin{subequations}
   \begin{align}
    \mathbf{h}_1 & = \mathbf{x} \, ,\\ 
    \mathbf{h}_{l+1} & = \mathrm{ReLU} \left(\mathbf{W}_{l} \mathbf{h}_{l} + \mathbf{b}_{l}\right), \, l = 1, 2,\, \ldots, L-2 \, , \\
    \mathbf{h}_{L} & := [\mathbf{h}_L^P;\mathbf{h}_L^V] = \mathbf{W}_{L-1} \mathbf{h}_{L-1} + \mathbf{b}_{L-1} , \label{lin_layer}\\
    \mathbf{V}_{gr}&= \mathrm{LinProj}(\tanh(\mathbf{h}_{L}^P)) \, ,\\
    \Delta\mathbf{P}_g^{\text{AC}} &= \mathbf{h}_L^V, \label{lin_layer_pg}\\ 
    \mathbf{P}_g &= \min\, (\max\, (\mathbf{\underline{P}}_g, \mathbf{P}_g^{\text{DC}} + \Delta\mathbf{P}_g^{\text{AC}}), \mathbf{\bar{P}}_g)\, . \label{pg_hard}
\end{align} 
\end{subequations}
where $\mathbf{V}_{gr}:= [\mathbf{V}_g; V_\text{ref}] \in \mathbb{R}^{N_g + 1}$. As shown in Eqs.~\eqref{lin_layer} and \eqref{lin_layer_pg}, a linear output layer for the $\Delta\mathbf{P}_g^{\text{AC}}$ is adopted because its value can be either positive or negative. A hard clip is used to ensure that $\mathbf{P}_g$ satisfies the box constraints. For the voltage magnitudes, the activation function $\tanh(\cdot)$ is used to generate bounded outputs, and a linear operator $\mathrm{LinProj}(\cdot)$ is further adopted to map within the box constraints.

As shown in Fig.\ref{framework}, given $\mathbf{y}$, the optimization model aims to find the closest optimal setpoints $\tilde{\mathbf{y}}:= [\tilde{\mathbf{P}}_g; \tilde{\mathbf{V}}_g; \tilde{V}_\text{ref}]$ that satisfies the PF and branch flow equations. The closest and optimality are achieved by minimizing the projection difference and the generation cost in the objective function, respectively. The advantages of introducing the optimization model are summarized as: 
\begin{itemize}
    \item The computational time of solving the optimization-based PF model under the NN guidance can be reduced compared to that of the original AC-OPF problem.
    
    \item The solution feasibility can be improved by integrating the hard constraints in the optimization model instead of adding a soft loss penalty to the training loss. 
    
    \item The projected gradient leverages the optimal Lagrange multipliers given the objective function to provide the best gradient descent direction for the NN updates. The soft loss typically utilizes fixed Lagrange multipliers, and while it provides a valid gradient descent direction, it is typically not the best gradient descent direction.
 
    \item Compared to the learning framework that learns the AC-OPF mapping directly, the proposed NN can obtain an AC-OPF solution that is close to the DC-OPF solution by learning the residual corrections. 
    
\end{itemize}

Let $\tilde{\mathbf{u}} := [\tilde{\mathbf{P}}_g, \tilde{\mathbf{Q}}_g, \tilde{P}_{g, \text{ref}}, \tilde{Q}_{g, \text{ref}}, \tilde{\mathbf{V}}, \tilde{\boldsymbol{\theta}}]$ collect all the decision variables. Instead of relying solely on the optimization solver to determine the optimal solution, the NN guides the search direction by providing a reference value for the objective function. The objective function consists of two parts: 1) the generation cost and 2) the projection loss, represented by an $\ell_2$-norm term. The optimization-based PF feasibility restoration model can be formulated as:
\begin{subequations}
\label{AC-PF1}
\begin{align} 
\min_{\tilde{\mathbf{u}}} \, L_o :=& w_p\sum_{i \in \mathcal{N}_g } (\tilde{P}_g - P_g)^2 + w_v\sum_{i \in \mathcal{N} \setminus \mathcal{N}_d} (\tilde{V}_i - V_i)^2  \\
    +& w_o\sum_{i} c_i(P_{g, i}) \notag\, ,
    \label{oj1}\\
    \textrm{s.t.} \quad \eqref{pf1}&-\eqref{pg} \, ,
\end{align} 
\end{subequations}
Note that the value of $\mathbf{y}$ is given by the NN, which implies $\mathbf{y}$ is considered to be a fixed parameter in problem \eqref{AC-PF1}. Removing the projection loss by setting the non-negative weighting parameters $w_p$ and $w_o$ to 0, the problem \eqref{AC-PF1} degrades to the original AC-OPF problem \eqref{AC-PF}. Compared to the original AC-OPF problem \eqref{AC-PF}, whose objective function involves only the generation cost, the proposed formulation \eqref{AC-PF1} introduces a multi-objective problem that may sacrifice optimality. Nevertheless, this drawback can be alleviated by assigning a significant positive value to the weighting parameter $w_o$.

\end{subsection}

\begin{subsection}{Training loss function}
Thanks to the optimization model that ensures feasibility and optimality, the training loss has a single objective, i.e., the difference between the NN output and the closest optimal solution. The training loss function $L$ is formulated as:
\begin{equation} 
    \label{Ld}
    L := \| \mathbf{P}_g - \mathbf{\tilde{P}}_g (\mathbf{y})\|_2 + w_v (\| \mathbf{V}_{gr} - \mathbf{\tilde{V}}_{gr}(\mathbf{y})\|_2 ) \, ,
\end{equation} 
where $\mathbf{V}_{gr} = [\mathbf{V}_g; V_\text{ref}] \in \mathbb{R}^{N_g + 1}$ collects the voltage magnitudes of the generator buses and the reference bus. $w_v$ is a positive weighting parameter to balance the relative magnitude of different loss terms. This is due to the operating range of the active power generation output being much more extended than the voltage magnitudes.
\end{subsection}

\begin{subsection}{Gradient computation}
In the backpropagation, the gradients of the training loss w.r.t. the NN weights should be calculated:
\begin{align}
\frac{\mathrm{d} L(\mathbf{y}, \tilde{\mathbf{u}}(\mathbf{y}))}{\mathrm{d} \mathcal{W}} &:= \left(\frac{\partial L}{\partial \mathbf{y}} + \frac{\partial L}{\partial \tilde{\mathbf{u}}} \times \frac{\mathrm{d} \tilde{\mathbf{u}}}{\mathrm{d} \mathbf{y}} \right)\times \frac{\mathrm{d}\mathbf{y}}{\mathrm{d} \mathcal{W}} \, ,  \label{loss}
\end{align}
where $\mathcal{W}$ collects the NN weights and bias matrices. The gradient computation of $\frac{\partial L}{\partial \mathbf{y}}$, $\frac{\partial L}{\partial \tilde{\mathbf{u}}}$, and $\frac{\mathrm{d}\mathbf{y}}{\mathrm{d} \mathcal{W}}$ are straightforward. This section focuses on the derivation of the implicit gradient computation $\frac{\mathrm{d} \tilde{\mathbf{u}}}{\mathrm{d} \mathbf{y}}$, which can be obtained based on the KKT conditions at the optimal solution. 

The optimization model \eqref{AC-PF1} can be formulated as a generic optimization problem: 
\begin{subequations}
\label{gen1}
\begin{align}
    \min_{\tilde{\mathbf{u}}} \quad & f(\tilde{\mathbf{u}};\mathbf{y}) \, , \\
    \textrm{s.t.} \quad & \mathbf{h}(\tilde{\mathbf{u}};\mathbf{y}) = \mathbf{0}  \, , \\
    & \mathbf{g}(\tilde{\mathbf{u}};\mathbf{y}) \leq \mathbf{0}. \label{ineq}
\end{align}
\end{subequations}
A slack variable $\mathbf{s}$ can be introduced to convert the inequality constraint \eqref{ineq} to an equality constraint. The optimization problem \eqref{gen1} can be rewritten as:
\begin{subequations}
\label{gen}
\begin{align}
    \min_{\tilde{\mathbf{u}}} \quad & f(\tilde{\mathbf{u}};\mathbf{y})  \, ,\\
    \textrm{s.t.} \quad & \mathbf{r}(\tilde{\mathbf{u}};\mathbf{y}) = \mathbf{0}  \label{eq}  \, , \\
    & \mathbf{s} \geq \mathbf{0} \label{ineq_dual}. 
\end{align}
\end{subequations}
Let $\boldsymbol{\lambda}$ and $\boldsymbol{\nu}$ represent the Lagrangian multipliers of the equality constraint \eqref{eq} and the inequality constraint \eqref{ineq_dual}, respectively. When the optimization problem is solved, the KKT conditions at the optimal primal and dual variables should be satisfied, including first-order optimality, primal feasibility, and complementary slackness. The KKT matrix is given in Eq.~\eqref{kkt} \cite{duvenaud2023, Pirnay2012}:
\begin{align} \label{kkt}
K(\tilde{\mathbf{u}}^*, \mathbf{s}^* , \boldsymbol{\lambda}^*, \boldsymbol{\nu}^*):= 
\begin{bmatrix}
\nabla_{\tilde{\mathbf{u}}} f(\tilde{\mathbf{u}}^*;\mathbf{y}) + \nabla_{\tilde{\mathbf{u}}} \mathbf{r}(\tilde{\mathbf{u}}^*;\mathbf{y})^\top \boldsymbol{\lambda}^*  \\
\mathbf{r}(\tilde{\mathbf{u}}^*;\mathbf{y}) \\
\mathbf{s}^* \circ \boldsymbol{\nu}^*
\end{bmatrix} = \mathbf{0} \, ,
\end{align}
where $\circ$ denotes the element wise product and the superscript $(\cdot)^*$ denotes the optimal value of the corresponding variable. By applying the implicit function theorem, the gradient of the optimal solution w.r.t. the parameter can be obtained by solving Eq.~\eqref{grad}~\cite{duvenaud2023, Pirnay2012, Jiayu2024}:
\begin{align}
&\frac{\mathrm{d} K\left(\tilde{\mathbf{u}}^*(\mathbf{y}), \mathbf{s}^*(\mathbf{y}) , \boldsymbol{\lambda}^*(\mathbf{y}), \boldsymbol{\nu}^*(\mathbf{y})\right)}{\mathrm{d} \mathbf{y}} \notag \\
& = \frac{\partial K^*}{\partial (\tilde{\mathbf{u}}, \tilde{\mathbf{s}}, \boldsymbol{\lambda}, \boldsymbol{\nu})}  \frac{\mathrm{d} (\tilde{\mathbf{u}}^*, \tilde{\mathbf{s}}^*, \boldsymbol{\lambda}^*, \boldsymbol{\nu}^*)}{\mathrm{d} \mathbf{y}} + \frac{\partial K^*}{\partial \mathbf{y}} = \mathbf{0}, \label{grad}.
\end{align}
Given that $\frac{\partial K^*}{\partial (\tilde{\mathbf{u}}, \tilde{\mathbf{s}}, \boldsymbol{\lambda}, \boldsymbol{\nu})}$ is non-singular \cite{Pirnay2012}, Eq.~\eqref{grad} can be rewritten as:
\begin{equation}
\frac{\mathrm{d} (\tilde{\mathbf{u}}^*, \tilde{\mathbf{s}}^*, \boldsymbol{\lambda}^*, \boldsymbol{\nu}^*)}{\mathrm{d} \mathbf{y}} = \left(\frac{\partial K^*}{\partial (\tilde{\mathbf{u}}, \tilde{\mathbf{s}}, \boldsymbol{\lambda}, \boldsymbol{\nu})}\right)^{-1} \frac{\partial K^*}{\partial \mathbf{y}}.
\label{backprop}
\end{equation}
After obtaining $\frac{\mathrm{d} \tilde{\mathbf{u}}^*}{\mathrm{d} \mathbf{y}}$, the gradient of the training loss function w.r.t. the NN output, i.e., $\frac{\mathrm{d} L}{\mathrm{d} \mathcal{W}}$ in Eq.~\eqref{loss}, can be further calculated for backpropagation. 
\end{subsection}

\begin{subsection}{Replay buffer}
Experience replay can improve the learning efficiency of the agent by leveraging past experiences and has been widely adopted in various reinforcement learning algorithms. Inspired by this, we introduce a replay buffer strategy to enhance the learning efficiency by reusing the past data pairs. The historical data pairs generated by the optimization model during each training epoch are stored in a buffer, denoted by $\mathcal{B} = \{ ((\mathbf{P}_d, \mathbf{Q}_d, \mathbf{P}_g^\text{DC}), (\mathbf{\tilde{P}}_g^s, \mathbf{\tilde{V}}_{gr}^s))\}$. The training loss can be expressed as:
\begin{equation} 
    \label{Ls}
    L_s := \| \mathbf{P}_g - \mathbf{\tilde{P}}_g^s\|_2 + w_v (\| \mathbf{V}_{gr} - \mathbf{\tilde{V}}_{gr}^s\|_2 ) \, ,
\end{equation} 
where $\mathbf{\tilde{P}}_g^s$ and $\mathbf{\tilde{V}}_{gr}^s$ are fixed targets, not as the functions of $\mathbf{y}$ in Eq.~\eqref{Ld}. Hence, the gradient computation $\frac{\mathrm{d} L_s(\mathbf{y})}{\mathrm{d} \mathcal{W}}$ is straightforward. The adopted supervised loss eliminates the need to compute projected gradients and reduces the overall training time. 

Here, the buffer is initialized at the beginning of each epoch, and $(\mathbf{\tilde{P}}_g^s, \mathbf{\tilde{V}}_{gr}^s)$ are generated based on the NN and the optimization model in the current training epoch. This distinguishes the replay buffer strategy from the supervised learning framework, because the latter has the offline-generated ground truth AC-OPF targets, which makes the NN learning process easier. However, as training progresses, the NN generates increasingly better solutions, which in turn leads to a gradual improvement in the label quality. To this end, our proposed learning scheme is listed as Algorithm \ref{alg}.

\begin{algorithm}[ht!]
\caption{Proposed learning scheme}\label{alg}
\begin{algorithmic}[1]
\Require Load demand and DC-OPF solution dataset. \;

\For{epoch $i = 1, 2, \ldots, n_o$}: \;
\State Sample load demand $\mathbf{x}$ and DC-OPF solution $\mathbf{P}_g^\text{DC}$. \;
\State Initialize the buffer $\mathcal{B}$.
\State Compute $\mathbf{y}$ through forward propagation. 
\State Obtain $\mathbf{\tilde{u}}$ by solving the optimization model Eq.~\eqref{AC-PF1}.
\State Calculate the training loss~\eqref{Ld} and update NN weights $\mathcal{W}$ in backpropagation based on the Eq.\eqref{backprop}.
\State Store $((\mathbf{x}, \mathbf{P}_g^\text{DC}), \mathbf{\tilde{y}})$ in $\mathcal{B}$. \;
\For{epoch $i = 1, 2, \ldots, n_i$} : \;
\State Sample $\mathbf{x}$ and $\mathbf{P}_g^\text{DC}$ from $\mathcal{B}$.
\State Compute $\mathbf{y}$ through forward propagation. 
\State Calculate the training loss~\eqref{Ls} and update $\mathcal{W}$.
\EndFor
\EndFor
\end{algorithmic}
\end{algorithm}

\end{subsection}

\subsection{Evaluation with PF Solver}
By incorporating the $\ell_2$ norm loss of the NN outputs in the objective function, the computation time of solving the proposed optimization model is faster than that of the original AC-OPF model. However, solving the optimization problem via \emph{Ipopt} requires more computation time than solving the PF equations using a PF solver. To further speed up the computation time, we replace the optimization model with the PF solver in the testing phase. Under limited decision time, the NN can be adopted to generate highly feasible and near-optimal solutions directly. Fig.~\ref{framework_pf} demonstrates the testing framework that incorporates a subsequent PF solver, where the NN is trained using the proposed hybrid framework. As a baseline, the capability of the NN trained by $M_{\text{proj}}$ in obtaining the PF solution is also tested, denoted by $M^\text{pf}_{\text{proj}}$.

\begin{figure}[ht!]
        \centering
    \includegraphics[width=1\linewidth]{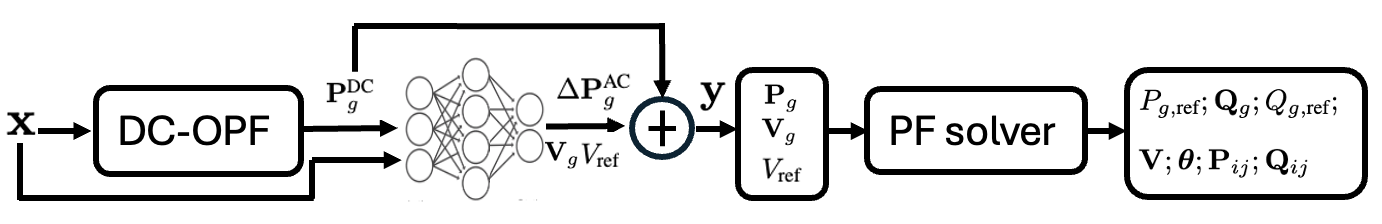}
    \caption{The testing framework with a subsequent PF solver $M^\text{pf}_{\text{prop}}$.}
    \label{framework_pf}
\end{figure}

\end{section}

\begin{section}{Numerical results}\label{sec:results}
This section demonstrates the effectiveness of the proposed framework on two benchmark systems: the small-scale IEEE 118-bus system and the large-scale PEGASE 9241-bus system.

\begin{subsection}{Simulation setup} 
Let $\mathbf{x}_{\text{nomi}}$ denote the nominal value of the load demand data given in \texttt{Matpower 8.0}. The load demand samples are generated randomly from a uniform distribution over the interval $[0.8\mathbf{x}_{\text{nomi}}, 1.2\mathbf{x}_{\text{nomi}}]$. The per-unit data is used for NN training, and the base MVA is 100MVA. The weighting parameters are set to $w_o = 10$,  $w_p = 1$, and $w_v = 100$ for the 9241-bus system, and to $w_o = 1$,  $w_p = 0.1$, and $w_v = 100$ for the IEEE-118 bus system during training. Due to the voltage magnitudes operating in a tight range, a significant value of $w_v$ helps mitigate the voltage magnitude estimation errors. The trade-off between optimality and feasibility is determined by the ratio of $w_o$ to $w_p$ and $w_v$. In the testing phase, the NN provides the reference values of the active power setup values by introducing projection loss in the objective function. This speeds up the computational time while sacrificing the optimality due to the multi-objective setup. Hence, we increase the weighting parameter to $w_o = 10^4$ and decrease $w_p = 0.1$ in the optimization model during testing to promote optimality. The optimization model in the learning framework is built based on Pyomo \cite{bynum2021pyomo}. 

The number of training, validation, and testing samples is $350$, $50$, and $100$, respectively. The FCNN has two hidden layers with $512$ neurons. The NN training is implemented on \texttt{Pytorch} using the Adam optimizer, with a mini-batch size of $32$. Considering the inherently data-driven nature of NN learning, it is essential to update the model frequently and promptly with new data instances. Therefore, we limit the model training time to one week to facilitate weekly updates based on the latest data. The learning rate schedule is adopted to enable faster convergence during the initial training epochs and fine-tuning in later epochs. Specifically, the learning rate is set to $0.0005$ for the first 50 epochs and then reduced to $0.0001$ thereafter. The number of training epochs $n_o$ is set to $60$ and $150$ for the IEEE-118 bus system and PEGASE-9241 bus system, respectively. The training epoch $n_i$ for the buffer replay is set to $10$. The NN training and testing experiments of the proposed methods were conducted on a MacBook Pro with an Apple M2 Pro chip with 32GB of RAM. 

\end{subsection}

\begin{subsection}{Methods for comparison and performance criteria} 
As detailed in Table \ref{Methods}, we compare the performance of the proposed method with some existing state-of-the-art methods. More details are given below:

\renewcommand{\arraystretch}{1.3}
\begin{table*}[tb!]
\caption{Methods for comparison.}
\label{Methods}
\centering
\begin{tabular}{|c|c|c|c|c|c}\hline 
Framework & Method & NN mapping & Optimization environment & Training loss  \\\hline 
Unsupervised & ${M}_{\text{dc3}}$ \cite{donti2021dc3} & \multirow{3}{*}{$\mathbf{x} \mapsto \mathbf{y} \mapsto $ PF solver } & PF \& branch & Operating constraint  \\
AC-aided (warm-started) & ${M}_{\text{dc3}}^{\text{AC}}$ \cite{WANG2022} & &  flow equations & violation loss \\
DC-aided (warm-started) & ${M}_{\text{dc3}}^{\text{DC}}$ & &  \eqref{pf1} - \eqref{bf3} &  and generation cost \\
\hline
Unsupervised & ${M}_{\text{proj}}$ \cite{Jiayu2024} & $\mathbf{x} \mapsto \mathbf{y}$ & Min projection loss  & Projection and generation loss \\
\hline
DC-aided unsupervised & ${M}_{\text{prop}}$ & $\mathbf{x}, \mathbf{P}_g^{\text{DC}} \mapsto \Delta\mathbf{P}_g^{\text{AC}}, \mathbf{V}_{gr} $ & Min projection and generation loss & Projection loss or supervised loss (buffer)  \\
\hline
Optimizer & \emph{Ipopt} \cite{IPOPT} & Not applicable & Solve AC-OPF Eq.\eqref{AC-PF} & Not applicable \\
\hline
\end{tabular}
\end{table*}
\renewcommand{\arraystretch}{1.0}

    \begin{itemize}
        \item ${M}_{\text{proj}}$ uses the optimization model that minimizes the projection loss while subject to all the operational constraints. The training loss function instead combines the generation cost and the projection loss. 
        
        \item With the randomly initialized NN in ${M}_{\text{dc3}}$, the subsequent PF solver may fail to find any solution for NN training. To address this, we use the DC-OPF solution to train the NN in a supervised fashion by utilizing the $\ell_2$ norm loss and then adopt the learning framework of ${M}_{\text{dc3}}$ based on this warm-up model, denoted by ${M}_{\text{dc3}}^{\text{DC}}$. However, ${M}_{\text{dc3}}$ and ${M}_{\text{dc3}}^{\text{DC}}$ does not work well on the large-scale bus system. Hence, we further replace all the 1 voltage magnitudes of the DC-OPF solution with the ground truth voltage magnitudes of the AC-OPF solution to warm-start the NN, denoted by ${M}_{\text{dc3}}^{\text{AC}}$. 
    
    \end{itemize}

Given the solution obtained by the PF solver, let $\mathbf{v} := [\mathbf{P}_g, P_{g, \text{ref}}, \mathbf{Q}_g, Q_{g, \text{ref}}, \mathbf{V}, \mathbf{S}^2]$ collect the decision variables with the operating limits. Let $\mathbf{P}_{gr} := [\mathbf{P}_g, P_{g, \text{ref}}]$ and $\mathbf{Q}_{gr} := [\mathbf{Q}_g, Q_{g, \text{ref}}]$ collect the active and reactive power outputs of the generators and the reference bus, respectively. The decision variables in $\mathbf{\tilde{u}}$ obtained by the optimization model strictly satisfy the operating limits. We evaluate the performance of the proposed method based on the following criteria:
\begin{itemize}
    \item Feasibility: The equality constraints are always satisfied in the optimization model. The constraint violation loss for the decision variable with the box constraint is defined as $l (\mathbf{v}_b) := [\mathbf{v_b - \overline{v}}]^{+} + [\mathbf{\underline{v} - v_b}]^{+}$, and let $\bar{l}$ and $\tilde{l}$ represent its maximum and the mean values, respectively.  
    
    \item Optimality: The optimality gap is defined as $C_{\text{gap}} := \frac{C - C_{\emph{Ipopt}}}{C_{\emph{Ipopt}}} \times 100\%$, where $C_{\emph{Ipopt}}$ and $C$ denote the total generation cost of $\emph{Ipopt}$ and the NN, respectively. 
 
    \item Computation time: The computational time of one data instance, denoted by $T_{\text{comp}}$. The speed-up ratio $T_{\text{ratio}}$ is defined as the computational time of the learning framework to that of $\emph{Ipopt}$. 

\end{itemize}

\end{subsection}

\begin{subsection}{Simulation results}
Tested on the IEEE-118 bus system and the PEGASE-9241 bus system, we present results demonstrating the training process, optimality and feasibility, the robustness when a power flow solver replaces the optimization model, and finally, the robustness under single-line topology changes. 

\subsubsection{Training process}
Fig.~\ref{fig:train_loss} shows the training loss curves of the proposed method ${M}_{\text{prop}}$ and the competing method ${M}_{\text{proj}}$ on the PEGASE-9241 bus system. The training loss of ${M}_{\text{proj}}$ consists of the generation cost and the projection loss. To have a fair loss value comparison as ${M}_{\text{prop}}$, we only show the projection loss part of ${M}_{\text{proj}}$. Despite both adopting an optimization-based learning framework, ${M}_{\text{prop}}$ achieves a much better convergence profile in estimating the optimal setup values of the generators. This demonstrates the advantages of learning the correction mapping as well as the replay buffer strategy.

\begin{figure}[ht!]
    \centering
    \includegraphics[width=1\linewidth]{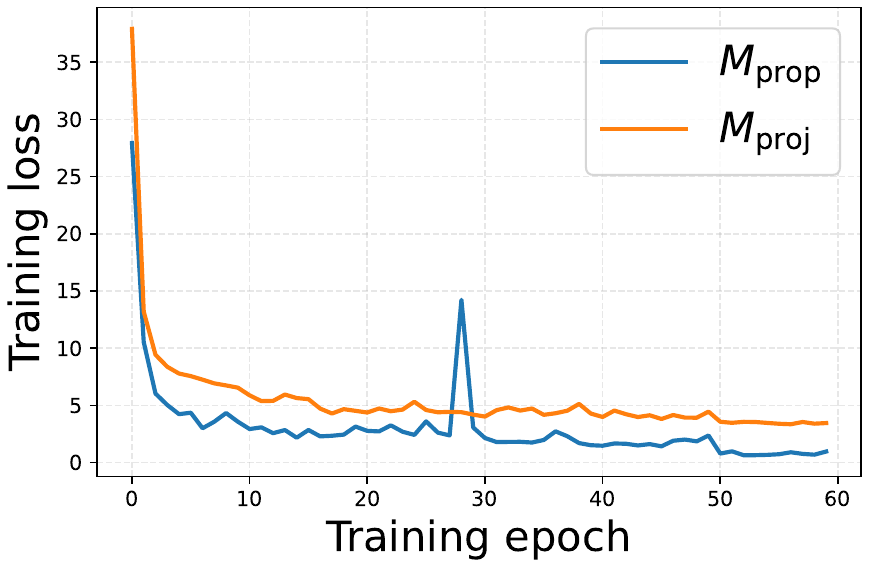}
    \caption{The evolution process of the training (projection) loss with the training epoch on the PEGASE-9241 bus system. The learning rate is reduced from $0.0005$ to $0.0001$ at epoch 50.}   
    \label{fig:train_loss}
\end{figure}

\subsubsection{Optimality and feasibility evaluation}
As shown in Table \ref{Testing}, the proposed method achieves strict satisfaction of all constraints in the AC-OPF problem through the embedded optimization environment. The proposed method achieves a smaller optimality gap than $M_\text{proj}$. The training loss function of ${M}_{\text{dc3}}^{\text{AC}}$ includes the constraint violation loss and generation cost. Increasing the relative importance of the constraint violation penalty term in the loss function contributes to less constraint violation. However, the soft loss design mechanism cannot avoid constraint violation. The training of ${M}_{\text{dc3}}$ fails because the subsequent PF solver cannot find any feasible solution given the randomly initialized NN. With the aid of the DC-OPF solutions for warm-started, the training of ${M}_{\text{dc3}}^\text{DC}$ still fails. Both lead to numerical issues in the power flow Jacobian computation, which demonstrates the advantage of introducing the optimization model to find the closest feasible solution.  

In addition, ${M}_{\text{dc3}}$ achieves promising performance on small bus systems with a limited number of decision variables. Table \ref{Testing3} shows the simulation results on the IEEE-118 bus system. The performance of the proposed method with ${M}_{\text{dc3}}$ is comparable. ${M}_{\text{dc3}}$ achieves a $0.11\%$ optimality gap with minor constraint violations. ${M}_{\text{dc3}}$ works well when the PF divergence does not occur for the small-scale bus system. Hence, the feasibility recovery layer in the proposed method is more preferable when the PF divergence occurs. 

\renewcommand{\arraystretch}{1.3}
\begin{table}[ht!]
\caption{Testing results on the PEGASE-9241 bus system. ``\checkmark" indicates that the loss can be avoided, which implies \emph{Ipopt} exits upon finding an optimal solution and the same notation applies to the other tables. ``\xmark" indicates the NN training fails due to numerical issues.}
\label{Testing}
\centering
\begin{tabular}{|c|c|c|c|c|c|c|}\hline 
Criteria  & ${M}_{\text{dc3}}$ & ${M}_{\text{dc3}}^\text{DC}$ & ${M}_{\text{dc3}}^{\text{AC}}$ & ${M}_{\text{proj}}$ & ${M}_{\text{prop}}$  \\
\hline
 $\bar{l}(\mathbf{v}) $ & \multirow{3}{*}{\xmark} & \multirow{3}{*}{\xmark} & 4.43  & \multirow{2}{*}{\checkmark} & \multirow{2}{*}{\checkmark}  \\ 
$\tilde{l}(\mathbf{v})$ & \xmark & & $6.5 \times 10^{-4}$ &  &  \\ 
$C_{\text{gap}}$ &  &  & 0.54\% & 0.72\% &  0.07\% \\

\hline
\end{tabular}
\end{table}
\renewcommand{\arraystretch}{1.0}

\renewcommand{\arraystretch}{1.3}
\begin{table}[ht!]
\caption{Testing results on the IEEE-118 bus system.}
\label{Testing3}
\centering
\begin{tabular}{|c|c|c|}\hline 
Criteria & ${M}_{\text{prop}}$ & ${M}_{\text{dc3}}$ \\
\hline
 $\bar{l}(\mathbf{v})$ & \checkmark & $8 \times 10^{-4} $  \\ 
$\tilde{l}(\mathbf{v}) $ & \checkmark &  $1 \times 10^{-6} $ \\ 
$C_{\text{gap}}$ & 0.00\% & 0.11\% \\
\hline
\end{tabular}
\end{table}
\renewcommand{\arraystretch}{1.0}

\subsubsection{Evaluation with PF Solver}
We demonstrate that the trained NN can effectively generate highly feasible and near-optimal solutions via the PF solver, without relying on the optimization model. This is appealing to real-time operation due to the significantly reduced computation time. As shown in Table \ref{decision_vio}, the proposed method decreases the constraint violation by $6.7\times$ compared with $M_{\text{dc3}}^\text{AC}$, while the optimality gap increases by only $1.1\times$. Note that the proposed method does not require any guidance from ground-truth data. The higher feasibility highlights the advantage of NN training with the projected gradient derived from the optimization model, compared to any valid gradient descent direction informed by the soft training loss. In addition, for the NN trained by $M_\text{proj}$, the constraint violation is significant, which indicates the NN trained by $M_\text{proj}$ cannot generate highly feasible solutions directly. This demonstrates the advantages of residual learning based on DC-OPF and the replay buffer strategy.

\renewcommand{\arraystretch}{1.3}
\begin{table}[hb!]
\caption{Averaged over the testing samples, the constraint violation of different decision variables and the optimality gap.}
\label{decision_vio}
\centering
\begin{tabular}{|c|c|c|c|c|c||c|}\hline
Method & $\bar{l}(\mathbf{P}_{gr})$ & $\bar{l}(\mathbf{Q}_{gr})$ & $\bar{l}(\mathbf{V})$ & $\bar{l}(\mathbf{S}_{ij}^2) $ & $\bar{l}(\mathbf{v})$ & $C_{\text{gap}}$ \\
\hline
${M}^\text{pf}_\text{prop}$ & 0.00 & 0.66 & 0.00 & 0.00 & \textbf{0.66} & 0.60\% \\
\hline
${M}_{\text{dc3}}^{\text{AC}}$  & 3.35 & 1.54 & 0.01 & 0.00 & 4.43 & 0.54\%  \\ 
\hline
${M}^\text{pf}_\text{proj}$  & 97 & 13 & 0.01 & 662 & 668 & 1.26\%  \\
\hline
\end{tabular}
\end{table}
\renewcommand{\arraystretch}{1.0}

The PEGASE-9241 bus system contains 1445 generators, and the constraint violations of ${M}_{\text{prop}}^\text{pf}$ primarily arise from reactive power generation limits of these generators. The closest feasible solution that satisfies all the constraints, denoted by $\mathbf{\breve{y}}$, can be obtained by solving the optimization problem Eq.~\eqref{AC-PF1} with $w_o = 0$. As shown in Fig.~\ref{fig:y_diff}, the estimation error between the NN output and the closest feasible solution is minimal. Evaluated over all testing samples, the mean estimation errors of $\mathbf{P}_g$ and $\mathbf{V}_{gr}$ are $3 \times 10^{-5}$ and $9 \times 10^{-6}$, respectively. Hence, the trained NN by ${M}_{\text{prop}}$ demonstrates promising capability in generating high-quality setpoints of the generators. However, the prediction errors are hard to completely avoid, even given the ground truth optimal AC-OPF solutions \cite{Fioretto2020}. Hence, the optimization model is necessary to ensure strict feasibility because the NN cannot predict the generator setup values without any errors. If the reactive power generation capacity violation is allowable to a certain degree, using the PF solver is more computationally efficient for real-time applications.

\begin{figure}[ht!]
    \centering
    \includegraphics[width=1\linewidth]{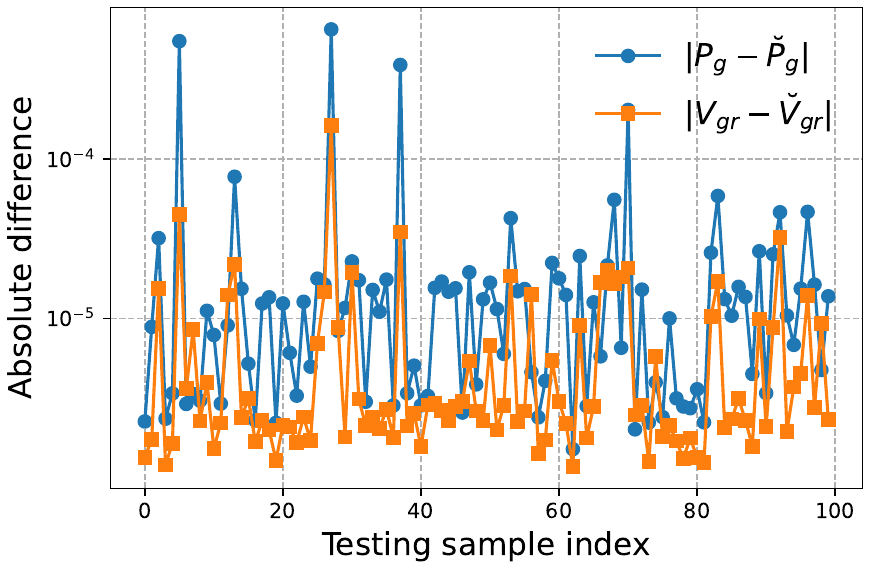}
    \caption{The average absolute difference of the NN output and the closest strict feasible solution of the testing samples on the PEGASE-9241 bus system.}   
    \label{fig:y_diff}
\end{figure}

As shown in Fig.~\ref{fig:diff_Pg_DC}, the generation setpoints in $\mathbf{P}_g$ obtained by the proposed method show the minimal deviation from the DC-OPF solution, with an average deviation less than half of $\mathbf{M}^\text{pf}_\text{proj}$. The generation setpoint $\mathbf{P}_g$ obtained by ${M}_{\text{dc3}}^{\text{AC}}$ is close to the DC-OPF solution because of the NN warmed-started training process. 

\begin{figure*}[ht!]
    \centering
    \includegraphics[width=1\linewidth]{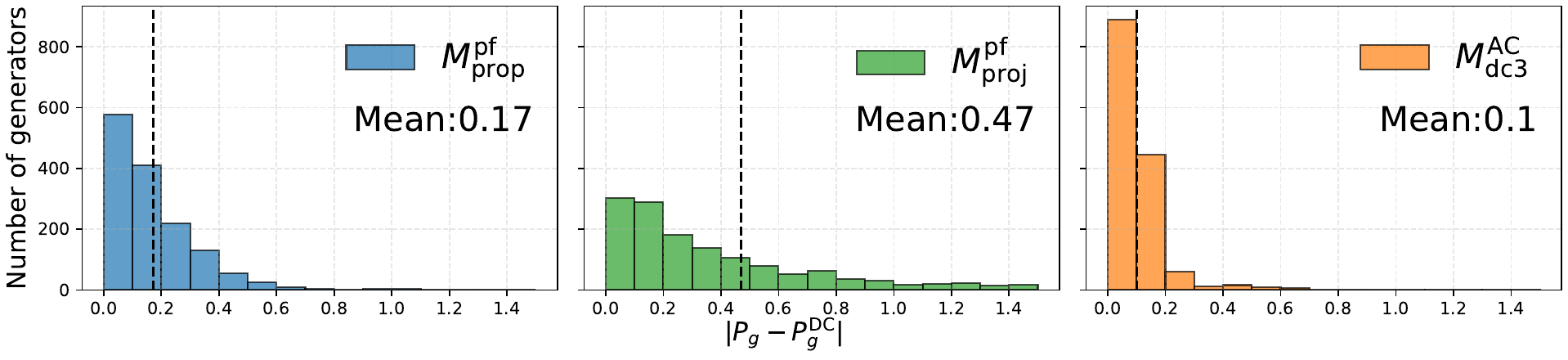}
    \caption{The histogram of the average power generation output difference of the NN output and the DC-OPF solution of the testing samples on the PEGASE-9241 bus system.} 
    \label{fig:diff_Pg_DC}
\end{figure*}

\subsubsection{Topology change}
We demonstrate the generalization ability of the proposed method under varying topologies. In the N-1 contingency analysis, the PF analysis needs to be repeatedly conducted for all the possible single-line outages. For example, the linearized branch flow estimation method, such as the line outage distribution factors, can be utilized to calculate the branch flows under the topology change. The proposed framework can rapidly compute branch flows without the need for initialization. The capability of the trained FCNN of $M_\text{prop}$ under topology change is explained as follows. The trained NN learns the correction relationship from the DC-OPF solution to the AC-OPF solution. When the line outage occurs, we remove that branch from the DC-OPF model, and this enables the DC-OPF model to capture the topology change. In this case, as long as the NN learns the correction relationship from DC-to-AC, it should achieve promising generalization ability against topology changes. 

To simulate the possible variations in topology, we randomly select and remove one branch from the power grid. Among all line contingency cases, only those with an optimal DC-OPF solution and a well-conditioned Jacobian are used for performance evaluation. The improvement of the DC-OPF model and PF solver is not considered in this work. 
\begin{itemize}
    \item PEGASE-9241 bus system: For one testing data instance, we randomly select 200 branches, collected in $\mathcal{M}_r \subset \mathcal{M}$. To consider a more challenging scenario, the branches are sorted in descending order according to the nominal values of the branch flows. Let $\mathcal{M}_l$ denote the set of the 200 branches with the largest branch flows. To demonstrate the trained NN's capability in learning the mapping from DC to AC under topology changes, we apply 400 line outage scenarios to the first 10 testing samples. The adopted testing contingency cases obtained through random selection and sorted selection are collected in $\mathcal{\hat{M}}_r \subset \mathcal{M}_r$ and $\mathcal{\hat{M}}_l \subset \mathcal{M}_l$, respectively. Specifically, we obtain $\frac{|\mathcal{\hat{M}}_r|}{|\mathcal{M}_r|} = 0.89$ and $\frac{|\mathcal{\hat{M}}_l|}{|\mathcal{M}_l|} = 0.87$. 

    \item IEEE-118 bus system: We simulate all possible single-line outage scenarios for all testing samples on this small-scale bus system. The adopted testing contingency cases are collected in $\mathcal{\hat{M}} \subset \mathcal{M}$ and $\frac{|\mathcal{\hat{M}}|}{|\mathcal{M}|} = 0.99$.
    
\end{itemize}

As shown in Table \ref{decision_vio_change}, the PF solver maintains promising performance under topology changes, exhibiting no significant degradation compared to the baseline cases without line outages.
    
\renewcommand{\arraystretch}{1.3}
\begin{table*}[ht!]
\caption{Under the line outage, the average constraint violation values of different variables using ${M}^\text{pf}_\text{prop}$. The constraint violation values without the line outage serve as the baseline.}
\label{decision_vio_change}
\centering
\begin{tabular}{|c|c|c|c|c|c|c|c|c}\hline
 Bus system & Set & $\bar{l}(\mathbf{P}_{gr})$ & $\bar{l}(\mathbf{Q}_{gr})$ & $\bar{l}(\mathbf{V}$ ) & $\bar{l}(\mathbf{S}_{ij}^2)$ & $\tilde{l}(\mathbf{v})$  \\
\hline
\multirow{3}{*}{PEGASE-9241} & $\mathcal{\hat{M}}_r$ & 0.00 & 0.64 & 0.00 & 0.00 & $2.1 \times 10^{-4} $\\
\cline{2-7}
& $\mathcal{\hat{M}}_l$ & 0.00 & 0.68 & 0.00 & 0.00 & $2.1 \times 10^{-4} $ \\
\cline{2-7}
& No line outage & 0.00 & 0.66 & 0.00 & 0.00 & $2.1 \times 10^{-4} $ \\
\hline 
\multirow{2}{*}{IEEE-118} & $\mathcal{\hat{M}}$ & 0.00 & 0.06 & 0.00 & 0.00 & $1.0 \times 10^{-4}$ \\
\cline{2-7}
& No line outage & 0.00 & 0.01 & 0.00 & 0.00 & $1.5 \times 10^{-5}$ \\
\hline 
\end{tabular}
\end{table*}
\renewcommand{\arraystretch}{1}

\subsubsection{Computation time comparison}
Table \ref{Testingtime} shows the average computational time of testing data. The proposed methods $M_{\text{prop}}$ and $M_{\text{prop}}^\text{pf}$ achieve $1.6\times$ and $40 \times$ speed up ratios compared to $\emph{Ipopt}$ on the PEGASE-9241 bus system, respectively. The computation time of ${M}_{\text{dc3}}^{\text{AC}}$ and $M^\text{pf}_{\text{prop}}$ is much less than ${M}_{\text{prop}}$, because solving the PF equations requires less time than solving an optimization problem. 

\renewcommand{\arraystretch}{1.3}
\begin{table}[ht!]
\caption{Computational time of one testing data instance on the PEGASE-9241 bus system}
\label{Testingtime}
\centering
\begin{tabular}{|c|c|c|c|c|c|c|}\hline 
Bus system & Time & \emph{Ipopt} & $M_{\text{prop}}$ & $M_{\text{proj}}$ & $M^\text{pf}_{\text{prop}}$ & ${M}_{\text{dc3}}^{\text{AC}}$  \\
\hline
\multirow{2}{*}{PEGASE-9241} & $T_{\text{comp}}$ & 8sec & 5sec & 5sec & 0.2sec & 0.2sec  \\
& $T_{\text{ratio}}$  & 1 & 1.6 & 1.6 & 40 & 40 \\
\hline
\end{tabular}
\end{table}
\renewcommand{\arraystretch}{1}

\end{subsection}

In summary, the proposed framework demonstrates a better training loss profile thanks to the replay buffer strategy, which enhances the training efficiency. The feasibility and optimality are ensured by adopting the optimization environment while reducing the computational time compared to solving the original AC-OPF problem. In addition, the trained NN can generate highly feasible PF solutions directly, even without the optimization model, and under the topology change. 

\end{section}

\begin{section}{Conclusion}\label{sec:conclusion}
This paper proposes an optimization-based learning framework for rapidly solving the AC-OPF problem based on the DC-OPF solution, enabling the independent system operator to obtain a near-optimal solution in real-time operation. The proposed framework does not rely on conventional optimization solvers to generate AC-OPF data pairs, thereby alleviating the time-consuming data preparation process. The optimization model is embedded in the learning framework to ensure feasibility, which effectively addresses the constraint violation issue compared to the soft loss penalty. Compared to conventional optimization solvers such as \emph{Ipopt}, the proposed method can achieve a significant speedup in solving time on the large-scale bus system.

In addition, instead of learning the optimal AC-OPF solution directly, the proposed NN learns the residual correction to the DC-OPF solution. This enables the network to generate active power setpoints that are close to the DC-OPF results. The replay buffer strategy is employed to reduce the NN training time by utilizing a supervised loss. The trained NN can rapidly generate highly feasible solutions directly without the NP-hard AC-OPF optimization model under perturbations in the network topology.

\end{section}

\bibliographystyle{IEEEtran}
\bibliography{ref}
\end{document}